\documentclass[preprint,3p,times]{elsarticle}

\usepackage{amsmath,amssymb,amsthm}
\usepackage{pgfplots} 
\pgfplotsset{compat=1.17} 
\usepackage{natbib}
\biboptions{authoryear}
\usepackage[hidelinks]{hyperref}

\newcommand{\HHH}{\mathbb{H}}
\newcommand{\III}{\mathbb{I}}
\newcommand{\HL}{H^*}
\newcommand{\IL}{I^*}
\newcommand{\DKL}{D}
\newcommand{\DJS}{{D^*}}

\newcommand{\subrel}[1]{\, #1 \,}

\theoremstyle{plain}
\newtheorem{theorem}{Theorem}[section]
\newtheorem{prop}[theorem]{Proposition}
\newtheorem{lemma}[theorem]{Lemma}
\newtheorem{corollary}[theorem]{Corollary}
\theoremstyle{definition}
\newtheorem{defn}[theorem]{Definition}
\newtheorem{proofpart}{Part}

\let\oldsqrt\sqrt
\def\sqrt{\mathpalette\DHLhksqrt}
\def\DHLhksqrt#1#2{%
\setbox0=\hbox{$#1\oldsqrt{#2\,}$}\dimen0=\ht0
\advance\dimen0-0.2\ht0
\setbox2=\hbox{\vrule height\ht0 depth -\dimen0}%
{\box0\lower0.4pt\box2}}
 
\begin{document}

\begin{frontmatter}

\title{An entropy functional bounded from above by one}
\ead{john.camkiran@utoronto.ca}
\author{John \c{C}amk{\i}ran}
\address{Department of Materials Science and Engineering, University of Toronto, Toronto, Ontario M5S 3E4, Canada}

\begin{abstract}
Shannon entropy is widely used to quantify the uncertainty of discrete random variables. But when normalized to the unit interval, as is often done in practice, it no longer conveys the alphabet sizes of the random variables under study. This work introduces an entropy functional based on Jensen-Shannon divergence that is naturally bounded from above by one. Unlike normalized Shannon entropy, this new functional is strictly increasing in alphabet size under uniformity and is thus well suited to the characterization of discrete random variables.
\end{abstract}


\end{frontmatter}

\section{Introduction}
\label{sec:intro}

Long since its introduction, Shannon entropy remains the standard way of quantifying the uncertainty of a discrete random variable \citep{cover_2012}. In practice, it is often normalized to the unit interval through division by its maximum value of $\log_2 N$, where $N$ denotes the number of symbols in the alphabet of the random variable \citep{studholme_1999, zhou_2014, antonelli_2017}. However, a key property of Shannon entropy is lost in its normalization, namely its strictly increasing monotonicity in $N$ for uniformly distributed random variables. The loss of this property is sometimes desirable, such as when studying discret\textit{ized} continuous random variables, wherein $N$ is subjectively chosen. Yet it is preclusive to the characterization of genuinely discrete random variables, to which $N$ is inherent. This is easily seen by comparing the fair coin toss and the fair dice roll, which are distinct in uncertainty but equal in normalized entropy. A natural question to ask, therefore, is whether it is possible for an entropy functional to be simultaneously bounded from above by one and strictly increasing in alphabet size under uniformity. The present work answers this question in the affirmative. Throughout the work, one-bounded is used to mean ``bounded from above by one'', and distribution refers to the probability mass function of a discrete random variable. Our main result is formalized in the following definition and theorem:

\begin{defn}[entropy functional]
\label{defn:entropy_functional}

Let $\mathcal{P}$ denote the set of all distributions over nonempty alphabets $\mathcal{X}$. An \textit{entropy functional} is a map $\HHH : \mathcal{P} \to \mathbf{R}$ with the following properties: \vspace{0.5\baselineskip}\newline
\renewcommand{\arraystretch}{1.1}
\begin{tabular}{rll}
    I. &  $\HHH(p) \geq 0$ & \textnormal{(nonnegativity)} \\
    II. &  $\lim_{p' \to p} \HHH(p') = \HHH(p)$  & \textnormal{(continuity)} \\
    III. &  $\HHH(p')=\HHH(p)$ for all $p'$ such that $p'(\mathcal{X}) = p(\mathcal{X})$  & \textnormal{(symmetry)} \\
    IV. &  $\HHH\left[ \lambda p +(1-\lambda) p\right] \geq \lambda \HHH(p) + (1-\lambda) \HHH(p)$ for all $\lambda \in [0,1]$ & \textnormal{(concavity)} \\
    V. & $\HHH(p) = \sum_{x} p(x) \III[p(x)]$ for some $\III$ that is nonnegative, decreasing, and strictly convex  & \textnormal{(expectation)} \\
    VI. & $\HHH(p) = 0$ if and only if  $p(x) = 1$ for some $x \in \mathcal{X}$ & (minimality) \\
    VII. & For $\mathcal{X}$ finite, $\HHH(p)$ is maximized if and only if $p$ is uniform & (maximality) \\
    VIII. & $\HHH( U_{N+1} ) > \HHH( U_N )$, where $U_N$ denotes the uniform distribution over an alphabet of size $N$ & (monotonicity)
\end{tabular}
\end{defn}

\begin{theorem}
There exists an entropy functional $\HHH$ such that $\HHH(p) \leq 1$.
\label{thm:one_bounded}
\end{theorem}

We proceed as follows: First, in \mbox{Section \ref{sec:definition}}, we define an information quantity that is bounded from above by one. Then, in \mbox{Section \ref{sec:proof}}, we show that its functional satisfies Definition \ref{defn:entropy_functional}, thereby proving Theorem \ref{thm:one_bounded}.

\section{Defining a one-bounded information quantity}
\label{sec:definition}

Given a distribution $p$ over a nonempty alphabet, the \textit{Shannon entropy} $H(p)$ is defined by
\begin{equation}
    H(p) = -\sum_{x} p(x) \log_2 p(x),
\end{equation}
taking $0 \log_2 0 = 0$, and quantifies the uncertainty in a random variable $X \sim p$ \citep{shannon_1948}. Extending this idea, the uncertainty in one random variable $X \sim p$ relative to a second random variable $Y \sim q$ with the same alphabet is quantified by the \textit{Kullback-Leibler divergence} (KL divergence) $\DKL \left(p \parallel q \right)$, defined by
\begin{equation}
    \DKL \left(p \parallel q \right) = \sum_{x} p(x) \log_2 \frac{p(x)}{q(x)}
\end{equation}
\citep{kullback_1951}.

These two quantities are in fact closely related. Define the \textit{self-joint distribution} $\delta$ as the joint distribution of $X$ with a \textit{deterministic} copy of itself,
\begin{equation}
    \delta(x,x') := 
    \begin{cases}
        p(x) & \text{if } x = x',    \\
        0    & \text{if } x \neq x'; \\
    \end{cases}
\end{equation}
and the \textit{self-product distribution} $\pi$ as the joint distribution of $X$ with an \textit{independent} copy of itself,
\begin{equation}
    \pi(x,x') := p(x)p(x').
\end{equation}
Then, the Shannon entropy of $p$ is equal to the KL divergence of $\delta$ with respect to $\pi$,
\begin{align}
    H(p) &= -\sum_{x} p(x) \log_2 p(x) \label{eq:align_ent} \\
    &= \phantom{-} \sum_{x} p(x) \log_2 \frac{p(x)}{p(x)^2} \\
    &= \phantom{-} \sum_{ x \subrel{=} x'} \delta(x, \, x') \log_2 \frac{\delta(x,x')}{\pi(x, \, x')} \\
     &= \phantom{-} \sum_{x \subrel{=} x'} \delta(x, \, x') \log_2 \frac{\delta(x, \, x')}{\pi(x,x')} + \underbrace{\sum_{x \subrel{\neq} x'} \delta(x, \, x') \log_2 \frac{\delta(x, \, x')}{\pi(x, \, x')}}_{\subrel{=} 0, \text{ since } \delta(x, \, x') \subrel{=} 0 \text{ for } x \subrel{\neq} x'} \\
    &= \phantom{-} \sum_{{x, \, x'}} \delta(x, \, x') \log_2 \frac{\delta(x, \, x')}{\pi(x, \, x')} \\
    &= \phantom{-} \DKL ( \delta \parallel \pi ). \label{eq:align_div}
\end{align}

The main insight of this work is that a one-bounded variant of Shannon entropy can be defined by adjusting the KL divergence in Equation (\ref{eq:align_div}) so that it is bounded from above by one. Such an adjustment to KL divergence is well known to exist and is called Jensen-Shannon divergence \citep{lin_1991}, which we now briefly review.

From Jensen's inequality and the concavity of $H$, for any two distributions $p$, $q$ over the same alphabet, we have
\begin{equation}
    H\left( \frac{p+q}{2} \right) \geq \frac{H(p)+H(q)}{2}.
\end{equation}
The \textit{Jensen-Shannon divergence} (JS divergence) $\DJS( p \parallel q)$ can be defined as the nonnegative difference between the left and right sides of this inequality,
\begin{equation}
    \DJS( p \parallel q) = H\left(\frac{p+q}{2}\right) - \frac{H(p)+H(q)}{2}.
    \label{eq:js_div_defn}
\end{equation}

JS divergence is known to be a one-bounded, symmetrized variant of KL divergence \citep{lin_1991}. The exact relationship between the two divergences can be revealed by expanding \mbox{Equation (\ref{eq:js_div_defn})} and letting ${m := (p+q)/2}$,
\begin{align}
    \DJS(p \parallel q) &= \frac{1}{2} \left[ - \sum_{x} p(x) \log_2 m(x) - \sum_{x} q(x) \log_2 m(x) + \sum_{x} p(x) \log_2 p(x) + \sum_{x} q(x) \log_2 q(x) \right] \\
    &= \frac{1}{2} \left[ \sum_{x} p(x)  \log_2 \frac{p(x)}{m(x)} + \sum_{x} q(x)  \log_2 \frac{q(x)}{m(x)} \right] \label{eq:js_div_expn} \\ 
    &= \frac{1}{2} \DKL(p \parallel m) + \frac{1}{2} \DKL(q \parallel m) .
\end{align}

Thus, the JS divergence $\DJS(\delta \parallel \pi)$ constitutes a one-bounded (and symmetric) quantity that otherwise behaves like the KL divergence ${\DKL ( \delta \parallel \pi )}$. But, by Equations (\ref{eq:align_ent}) to (\ref{eq:align_div}), the KL divergence ${\DKL ( \delta \parallel \pi )}$ is precisely the Shannon entropy $H(p)$. So, the JS divergence $\DJS(\delta \parallel \pi)$ must also constitute a one-bounded variant of Shannon entropy; we term it ``Lin entropy'' after the author of the seminal work \cite{lin_1991}.

\begin{defn}[Lin entropy] Given the marginal distribution $p$, self-joint distribution $\delta$, and self-product distribution $\pi$, of a discrete random variable, the \textit{Lin entropy} $\HL(p)$ is defined by
\[\HL(p) = \DJS ( \delta \parallel \pi ).\]
\label{defn:lin_entropy}
\end{defn}

\section{Proving the theorem}
\label{sec:proof}

Let us now prove \mbox{Theorem \ref{thm:one_bounded}} by showing that the one-bounded functional $\HL$ is an entropy functional in the sense of \mbox{Definition \ref{defn:entropy_functional}}. Properties I--III follow immediately from the properties of JS divergence \citep{lin_1991}. The rest of this section therefore focuses on Properties IV--VIII. 

Definition \ref{defn:lin_entropy} defines Lin entropy implicitly in terms of the self-joint and self-product distributions, making its analysis difficult. To facilitate the proof of Theorem \ref{thm:one_bounded}, the following proposition gives $\HL(p)$ explicitly in terms of the marginal distribution $p$:

\begin{prop}[Explicit form]
\[
    \HL(p) = \sum_{x} p(x)  \log_2 \sqrt{ \frac{4 p(x)^{p(x)}}{[p(x)+1]^{p(x)+1}} }.
\]
\begin{proof}
Using Definition \ref{defn:lin_entropy} and Equation (\ref{eq:js_div_expn}), we write
\begin{equation}
    \HL(p) = \frac{1}{2} \sum_{x, \, x'} \Bigg[ \delta(x,x') \log_2 \frac{\delta(x,x')}{m(x, \, x')} 
    + \pi(x,x') \log_2 \frac{\pi(x,x')}{m(x, \, x')} \Bigg].
\end{equation}
We split this summation into two terms based on the state of equality between $x$ and $x'$. For the first term we take $x = x'$, giving us $\delta(x, \, x') = p(x)$ and $\pi(x,x') = p(x)^2$ and thus
\begin{align}
    \label{eq:part_1}
    \HL_1(p) &= \frac{1}{2} \sum_{x} \Bigg[ p(x) \log_2 \left( \frac{2p(x)}{p(x)+p(x)^2} \right)
    + p(x)^2 \log_2 \left( \frac{2p(x)^2}{p(x)+p(x)^2} \right) \Bigg] \\
    &= \frac{1}{2} \sum_{x} p(x) \log_2 \left( \frac{2^{p(x)+1} p(x)^{p(x)}}{[p(x)+1]^{p(x)+1}} \right).
\end{align}
For the second term we have $x \neq x'$, giving us $\delta(x, \, x') = 0$ and $\pi(x,x') = p(x)p(x')$ and thus
\begin{align}
    \HL_2(p) &= \frac{1}{2} \sum_{x \subrel{\neq} x'} p(x) p(x') \log_2 \left( \frac{2p(x)p(x')}{p(x)p(x')} \right) \\ \label{eq:part_2}
    &= \frac{1}{2} \sum_{x \subrel{\neq} x'} p(x) p(x') \\
    &= \frac{1}{2} \left( \sum_{x, \, x'} p(x) p(x') -  \sum_{x \subrel{=} x'} p(x) p(x') \right) \\
    &= \frac{1}{2} \left( \sum_{x} p(x)  \sum_{x}{p(x)}  - \sum_{x} p(x)^2 \right) \\
    &= \frac{1}{2} \sum_{x} p(x) \log_2 \left( \frac{2}{2^{p(x)}} \right).
\end{align}
The proposition is now immediate from the sum of $\HL_1(p)$ and $\HL_2(p)$.
\end{proof}
\label{prop:explicit}
\end{prop}

\begin{figure}[t]
\centering
    \begin{tikzpicture}[baseline]
    	\begin{axis}[
    		title = (a),
    		ylabel = $\HL(U_N)$,
    		xlabel= $1/N$,
    		xmin = 0,
    		xmax = 1,
    		ymin = 0,
    		ymax = 1,
    		samples = 1000,
    		width=0.35\linewidth
            ]
            \addplot[thick,
                color=black
            ] {0.5*log2(4*x^x/((x+1)^(x+1)))};
    	\end{axis}
    \end{tikzpicture} 
\qquad \quad
    \begin{tikzpicture}[baseline]
    	\begin{axis}[
    		title = (b),
    		ylabel = ${\HL \big[\mathrm{Ber}(\alpha)\big]}$,
    		xlabel= $\alpha$,
    		xmin = 0,
    		xmax = 1,
    		ymin = 0,
    		ymax = 0.25,
    		samples = 1000,
    		width=0.35\linewidth
    		]
            \addplot[thick,
                color=black
            ] {0.5*(x*log2((4*x^x)/(x+1)^(x+1) )+0.5*((1-x)*log2((4*(1-x)^(1-x))/ (2-x)^(2-x))};
    	\end{axis}
    \end{tikzpicture}
\caption{$\HL$ for (a) a uniform random variable vs its reciprocal alphabet size $1/N$ and (b) a Bernoulli random variable vs its success probability $\alpha$.}
\label{fig:entropy}
\end{figure}
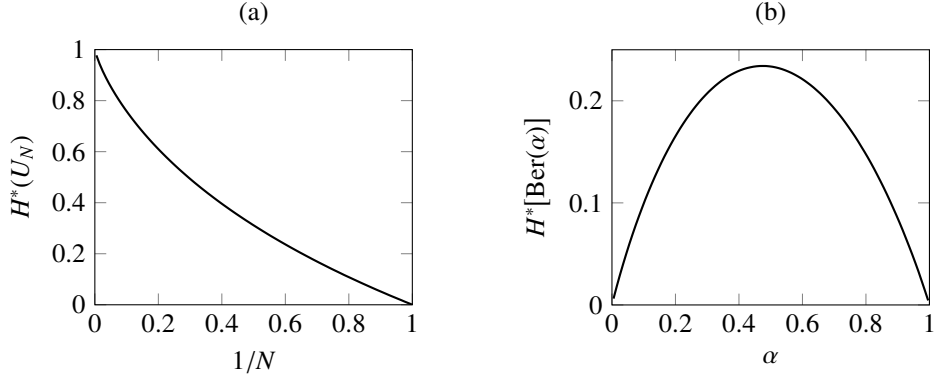

Before proceeding, we state an immediate consequence of Equations (\ref{eq:part_1}) and (\ref{eq:part_2}) that relates Lin entropy to an existing one-bounded information quantity called \textit{logical entropy} $h(p)$, defined by
\begin{equation}
    h(p) = \sum_{x \subrel{\neq} x'} p(x) p(x')
\end{equation} 
\citep{ellerman_2021}. The logical entropy functional satisfies every property from Definition \ref{defn:entropy_functional} except the strict convexity requirement in \mbox{Property V}. The significance of this requirement is that it captures the behaviour of the additive inverse of the logarithm, whose expectation is known to uniquely satisfy the Shannon-Khinchin axioms \citep{shannon_1948, khinchin_1957}.

\begin{corollary} 
$\HL(p) = \DJS \left( p \parallel p^2 \right) + h(p)/2.$
\end{corollary}

From its explicit form, it is easy to see that $\HL$ is strictly increasing in alphabet size under uniformity. This property is illustrated in Figure \ref{fig:entropy}(a) and formalized in the following lemma:

\begin{lemma}[Property VIII]
    Let $U_N$ denote the uniform distribution over a nonempty alphabet of size $N$. Then,
    \[\HL( U_{N+1} ) > \HL( U_{N} ) .\]
    \begin{proof}
    \begin{equation*}
    \HL(U_N) = \frac{1}{2} \log_2 \left( \frac{4 {(1/N)}^{1/N}}{(1/N+1)^{1/N+1}} \right),
    \end{equation*}
    which is strictly decreasing in $1/N$ and therefore strictly increasing in $N$.
    \end{proof}
\end{lemma}

The explicit form also makes it evident that Lin entropy can be thought of as the expectation of some quantity, which we term ``Lin surprisal'' in analogy with Shannon suprisal, the corresponding quantity for Shannon entropy. It can be shown that the Lin surprisal function is nonnegative, decreasing, and strictly convex.

\begin{defn}[Lin surprisal] Given the probability $p(x)$ of observing a symbol $x$, the \textit{Lin surprisal} $\IL\left[p(x)\right]$ is defined by
    \[\IL\left[p(x)\right] = \log_2 \sqrt{ \frac{4 p(x)^{p(x)}}{[p(x)+1]^{p(x)+1}} }. \]
\label{defn:lin_info}
\end{defn}

\begin{lemma}[Property V] \leavevmode
\begin{enumerate}[(i)]
    \item $\HL(p) = \sum_{x} p(x) \IL[p(x)]$.
    \item $\IL$ is nonnegative.
    \item $\IL$ is decreasing.
    \item $\IL$ is strictly convex.
\end{enumerate}
\begin{proof}
\begin{proofpart}
Statement (i) is visible from Proposition \ref{prop:explicit} and Definition \ref{defn:lin_info}.
\end{proofpart}
\begin{proofpart}
Statement (ii) is readily apparent given the inequality
\begin{equation*}
     \frac{4 p(x)^{p(x)}}{[{p(x)}+1]^{{p(x)}+1} } \geq 1,
\end{equation*}
which follows from the observation that the left-side ratio is strictly decreasing in $p(x)$ and equal to one for $p(x) = 1$.
\end{proofpart}
\begin{proofpart}
Statements (iii) and (iv) are immediate from the following two facts, respectively:
    \begin{alignat*}{3}
    &\frac{d \IL}{d{p(x)}} \null &&= \frac{\ln[{p(x)}]-\ln[{p(x)}+1]}{\ln(4)} & &< 0, \\[5pt]
    &\frac{d^2 \IL}{d{p(x)}^2} \null &&= \frac{1}{\left[{p(x)}^2+{p(x)}\right]\ln(4)} & &> 0.
\end{alignat*}
\end{proofpart}
\end{proof}
\label{lemma:expectation}
\end{lemma}

Much like its Shannonian counterpart, the Lin entropy functional $\HL$ is concave, and strictly so. This property is illustrated in \mbox{Figure \ref{fig:entropy}(b)} and formalized in the following lemma:

\begin{lemma}[Property IV] $\HL$ is (strictly) concave.
\begin{proof}
Since the sum of (strictly) concave functions is also (strictly) concave, it is enough to show that the summands of $\HL$ are (strictly) concave. Let $\alpha = p(x)$. Then,
\begin{equation}
    \frac{d^2 \left[\alpha \IL(\alpha)\right]}{d\alpha^2} = \frac{2(\alpha+1)\ln[\alpha/(\alpha+1)]+1}{(\alpha+1)\ln(2)} < 0,
    \label{eq:strict_concavity}
\end{equation}
where the inequality follows from the readily apparent fact that, since $\alpha$ is a probability, the numerator is always negative and the denominator always positive.
\end{proof}
\label{lemma:concavity}
\end{lemma}
 
The remaining two lemmas concern the extreme values of $\HL$, which occur as required by Properties VI and VII.

\begin{lemma}[Property VI]
Let $p$ be a distribution over a nonempty alphabet $\mathcal{X}$. Then,
\[ \HL(p) = 0 \quad \text{if and only if} \quad p(x) = 1 \quad \text{ for some }  x \in \mathcal{X}.\]
\begin{proof}
By its definition, $\HL(p) = \DJS(\delta \mid \pi)$. And from a basic property of JS divergence, $\DJS(\delta \mid \pi) = 0$ if and only if $\delta = \pi$ \citep{lin_1991}. To show that this occurs only when $p(x) = 1$ for some $x \in \mathcal{X}$, we consider the matrix representations of the various distributions involved: $p$ can be represented by a column matrix $P$, \mbox{$\delta$ by} the diagonal matrix $\mathrm{diag}(P)$, and $\pi$ the product ${P} {P}^\mathrm{T}$. Now the lemma is immediate from the observation that $PP^\mathrm{T} = \mathrm{diag}(P)$ if and only if $P$ has a single nonzero element, which must be equal to one by the unit measure axiom of probability.
\end{proof}
\label{lemma:min}
\end{lemma}

\begin{lemma}[Property VII] Let $p$ be a distribution over a finite nonempty alphabet. Then, $\HL(p)$ is maximized if and only if $p$ is uniform.
\begin{proof}
This proof is essentially identical to one of the proofs of an equivalent result for Shannon entropy \citep{conrad_2011}. By the continuity of $\HL$ and the compactness of its domain $\mathcal{P}$, we know that $\HL$ is maximized by some distribution. To prove that such a distribution must be uniform, it suffices to show that it cannot be nonuniform. We do so by demonstrating that if $p$ is nonuniform, there is always a distribution $q$ for which ${\HL(q) > \HL(p)}$.

Let $p(x), p(y)$ be two probabilities such that $p(x) < p(y)$, $\epsilon$ be a positive number such that ${p(x) + \epsilon < p(y) - \epsilon}$, and
\begin{equation}
    q(z) = 
    \begin{cases}
     p(z) + \epsilon & \text{if } z = x \\
     p(z) - \epsilon & \text{if } z = y \\
     p(z) & \text{otherwise}.
    \end{cases}
\end{equation}
Letting $f(\alpha) := \alpha \IL(\alpha)$, we have
\begin{align}
    \HL(q)  - \HL(p) &= f[q(x)] + f[q(y)] - f[p(x)] - f[p(y) \\
    &= \underbrace{f[p(x+\epsilon)] - f[p(x)]}_a  + \underbrace{\Big( f[p(y-\epsilon)] - f[p(y)] \Big)}_b  \\
    &> 0,
\end{align}
where the positivity follows from the fact, since $f$ is strictly concave (Equation \ref{eq:strict_concavity}), the sum of the positive parts of $a$ and $b$ is always greater than the sum of their negative parts.
\end{proof}
\label{lemma:max}
\end{lemma}

\section{Concluding remarks}
\label{sec:conc}

The present work introduces an entropy functional that is bounded from above by one. This functional provides a bounded characterization of discrete random variables that otherwise closely resembles Shannon entropy, as is often sought in practice. It is possible to extend the idea behind this work to define bounded analogues for other classical information quantities, such as mutual information and differential entropy.

\bibliographystyle{elsarticle-harv.bst}
\bibliography{main.bib}

\end{document}